\documentclass{iaesarticle2}

\usepackage{amsmath, amsfonts, amssymb, float, fancyhdr}
\usepackage[figuresright]{rotating}
\usepackage{authblk, graphicx, indentfirst, lastpage, lipsum}
\usepackage{pifont}

\makeatletter
\renewcommand{\@biblabel}[1]{[#1]\hfill}
\makeatother
\usepackage{subfig, caption, epstopdf}
\usepackage[left=3cm, right=2.5cm, top=2.5cm, bottom=2.5cm, includehead, includefoot]{geometry}
\usepackage[justification=centering]{caption}
\usepackage{titlesec}
\usepackage{xcolor}
\titleformat{\section}
{\normalfont\normalsize\bfseries\uppercase}{\thesection}{1.7em}{}
\titleformat{\subsection}
{\normalfont\normalsize\bfseries}{\thesubsection}{.95em}{}
\titleformat{\subsubsection}
{\bfseries}{\thesubsubsection}{.2em}{}
\titlespacing*{\section}{0cm}{0.7cm}{0cm}
\usepackage{enumitem}
\usepackage{appendix}

\usepackage[sorting=none]{biblatex}
\bibliography{Towards_ArXiv_v2.bib}

\AtEveryBibitem{%
  \clearfield{issn} 
  \clearfield{doi} 
  \clearfield{url}

}

\CopyrightLine[]{}{\textit{This is an open access article under the \textcolor{blue}{\underline{CC BY-SA}} license.}
\vspace{.5em}}

\author[1]{\bfseries Ngurah Indra Er}
\author[2,3]{\bfseries Kamal Deep Singh}
\author[4,5]{\bfseries Christophe Couturier}
\author[4,5]{\bfseries Jean-Marie Bonnin}

\affil[1]{Udayana University, Bali, Indonesia}
\affil[2]{Laboratoire Hubert Curien, Saint-Etienne, France}
\affil[3]{Universite Jean Monnet, Saint-Etienne, France}
\affil[4]{IRISA / IMT Atlantique, Rennes, France}
\affil[5]{Inria, Rennes, France}

\title{Towards A Simple and Efficient VDTN Routing Protocol for Data Collection in Smart Cities}
\shorttitle{Towards A Simple and Efficient VDTN Routing Protocol.. (Ngurah Indra Er)}

\begin{document}
\setcounter{page}{1}

\setlength{\parindent}{1.27cm}

\pagestyle{fancy}
\fancyhfoffset{0cm}

\journalname{International Journal of Electrical and Computer Engineering (IJECE)}
\journalshortname{Int J Elec \& Comp Eng}
\journalhomepage{http://ijece.iaescore.com}
\vol{x}
\no{x}
\months{August}
\years{202x}
\issn{2088-8708}
\DOI{10.11591}
\pagefirst{xx}
\pagelast{xx}

\maketitle

\hrule
\vspace{.1em}
\hrule
\vspace{.5em}
\noindent
\parbox[t][][s]{0.275\textwidth}{%
\textbf{Article Info}
\vspace{.5em}
\hrule
\vspace{.5em}
\begin{history}
\vspace{.5em}

Received mm dd, yyyy

Revised mm dd, yyyy

Accepted mm dd, yyyy

\vspace{.7em}
\end{history}

\vspace{.5em}
\hrule
\vspace{.5em}
\begin{keyword} 
\vspace{.5em}
Smart City \sep Data Collection \sep VDTN \sep ITS \sep V2X \sep Routing Protocol
\vspace{.5em}
\end{keyword}
\vspace{\fill}
}
\parbox{0.025\textwidth}{\hspace{0.5em}}
\parbox[t][][s]{0.7\textwidth}{%
\begin{abstract}
\vspace{.3em}
Smart cities today can utilize Vehicular Delay Tolerant Networks (VDTN) to collect data from connected-objects in the environment for various delay-tolerant applications. They can take advantage of the available Intelligent Transportation Systems (ITS) infrastructures to deliver data to the central server. The system can also exploit multiple and diverse mobility patterns found in cities, such as privately owned cars, taxis, public buses, and trams, along with their Vehicle-to-Everything (V2X) communications capabilities. In the envisioned convergence between the ITS and V2X, we believe that a simple and efficient routing protocol can be deployed for the delay-tolerant data delivery, contrary to the implementation of optimized solutions that might be resource-demanding and difficult to standardize. In this paper, we analyzed the performances of four baseline VDTN routing protocols, namely: Direct Delivery, First Contact, Epidemic, and Spray \& Wait, to understand their strengths and weaknesses. Our simulation results highlighted the trade-off between distinct approaches used by those protocols and pointed out some gaps that can be refined. This study provides new interesting ideas and arguments towards developing a simple, efficient, and high-performing routing protocol for data collection in smart cities.
\end{abstract}
}
\parbox[l]{\textwidth}{%
\rule{0.275\textwidth}{0.5pt} \hspace{0.5cm} \hrulefill
\\
\emph{\textbf{Corresponding Author:}}
\vspace{.5em}\\
Ngurah Indra Er,\\
Department of Electrical and Computer Engineering,\\
Udayana University,\\
Kampus Bukit Jimbaran, Badung, Bali, Indonesia.\\
Email: indra@unud.ac.id
}
\vspace{.5em}
\hrule
\vspace{.1em}
\hrule


\section{Introduction}
\label{}
The current and future smart cities will require data collection solutions from various connected-objects in their area. This in turn will help cities to manage natural resources intelligently, to ensure sustainable socio-economic development, and ultimately to enhance quality-of-life. The data collection's primary trend is to connect all sensors to a long-range operated network such as the cellular network and the Low-Power Wide Area Network (LPWAN). But as the number of sensors and the amount of data generated grows exponentially, the utilization of a cellular network to collect all kinds of data might not be economically feasible. At the same time, the use of LPWAN might be impeded by insufficient bandwidth. Therefore, it is desirable to keep such networks for only collecting data which has delay constraints. Whereas, other economical solutions should be utilized to collect delay-tolerant data.

Simultaneously, advances in vehicular technology have enabled the vehicle-to-vehicle (V2V), vehicle-to-infrastructure (V2I), and vehicle-to-everything (V2X) communications to connected-objects in their surroundings. Vehicles with current and future radio access networks will also have the capability to connect to the Internet. Furthermore, the regulatory body will soon make such communication capabilities compulsory for vehicles to support the safety-related applications \cite{machardy_v2x_2018}. This exciting new development will pave the way for vehicles to participate actively in the ecosystem of Smart Cities.

In terms of networking, Delay Tolerant Networks (DTN) enable communication even when there are connectivity issues. Such issues could be sparse and intermittent connectivity, long and variable delay, high latency, high error rates, highly asymmetric data rate, and even no end-to-end connectivity \cite{pereira2012delay}. The extension of Vehicular Ad hoc Networks (VANET) with DTN capabilities naturally leads to the concept of Vehicular Delay Tolerant Networks (VDTN), where the vehicular networks will be able to cope with the inherent intermittent connectivity during data delivery.

\begin{figure}[h!]
\centering
\includegraphics[width=0.8\linewidth]{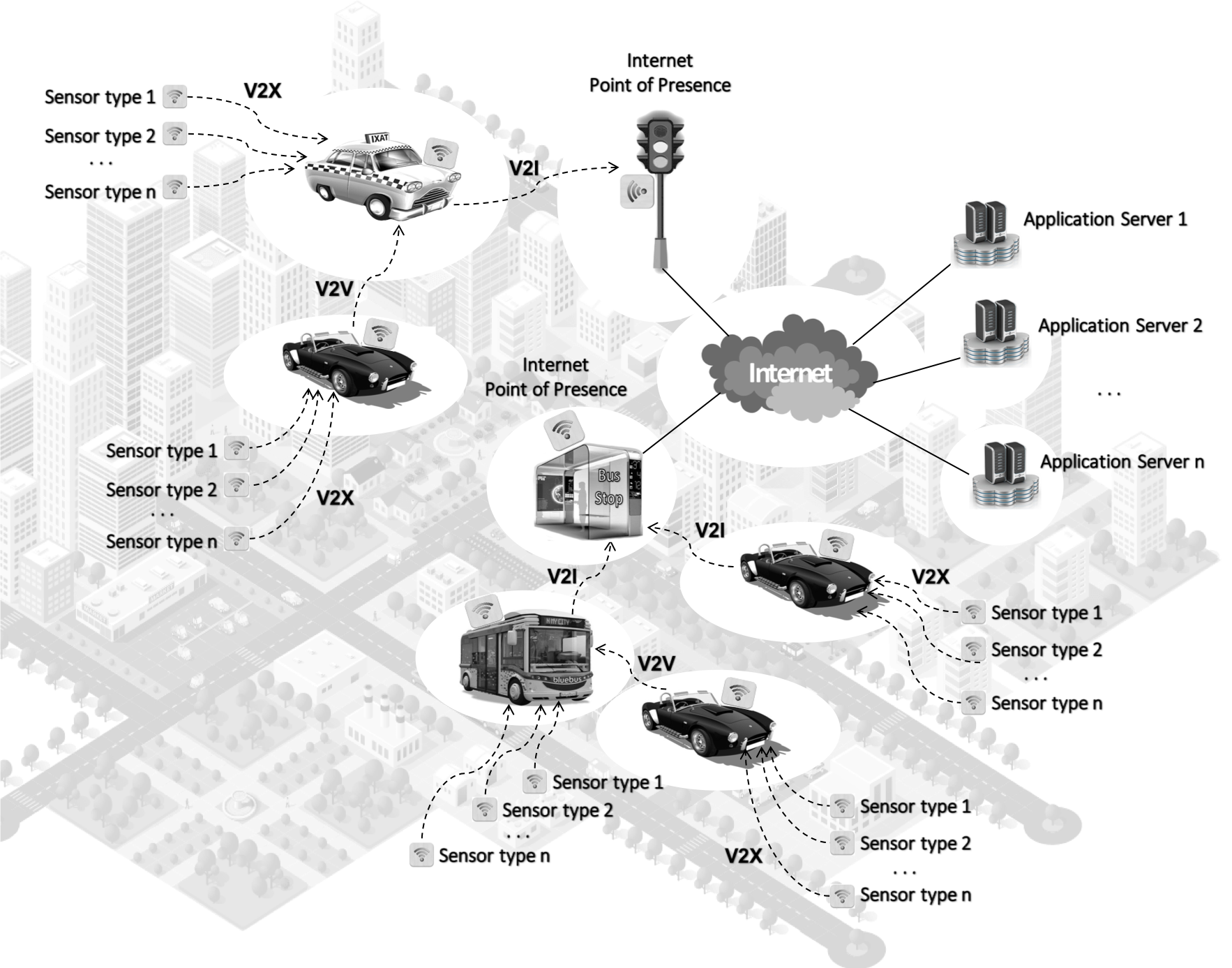}
\vspace{.7em}
\caption{The VDTN-Based Data Collection Scheme in Smart Cities}
\label{fig1}
\end{figure}

The Intelligent Transportation Systems (ITS), on the other hand, are advanced applications aiming to provide innovative services related to different modes of transport and traffic management, through vehicular communication, to improve road safety and comfort for drivers and passengers \cite{benamar_routing_2014}. Vehicles equipped with wireless devices can collect data from the environment and exchange traffic and road safety information with nearby vehicles, roadside units, and other connected objects. These functionalities are known as the vehicle-to-vehicle (V2V), vehicle-to-infrastructure (V2I), and the vehicle-to-everything (V2X) communications \cite{arena2019overview}. Concerning the implementation of VDTN in the ITS environment, we can expect an increasing number of ITS infrastructures available to assist in exchanging data within the networks. One crucial infrastructure element is the Road Side Unit (RSU), which can connect to the internet. Such units will be installed throughout the city, and they can function as a Point-of-Presence (PoP) for accessing the internet and forwarding data from vehicles to the core network. In another scenario that we do not consider in this paper, the RSU might not have an internet connection, but it can still participate in data forwarding as long as it has some wireless communication capability.

The PoPs will be available in numbers and they will be strategically placed, such as at traffic lights, road intersections, bus stops, and road lighting posts. Therefore, there will be several locations in the city where vehicles can offload the data (which they carry from connected-objects) to such PoPs, instead of having only one specific destination. Consequently, this unique configuration would need an efficient data routing approach. This makes the routing mechanism an integral part of the data collection process, which can crucially determine the data collection performance. Therefore, routing needs to be studied extensively to make sure its suitability for the chosen implementation. Before implementing the VDTN data collection system in the smart city environment, as depicted in Figure \ref{fig1}, we need to recognize and analyze the specific nature of the data forwarding and routing process and assess the performance of some baseline VDTN routing protocols. Their performances over defined benchmarks can then be examined to develop a better routing protocol for this specific purpose.

In this paper, we investigate the networking performance of four baseline VDTN routing protocols, namely: \emph{Direct Delivery}, \emph{First Contact}, \emph{Epidemic}, and \emph{Spray \& Wait}. These routing protocols implement the main routing concepts used in VDTN. We categorized them as `baseline' based on the fact that they use only a few parameters in their data forwarding decision. While some other optimized routing solutions are available in the literature, the study of these baseline protocols will give us the main trends. Interestingly, these four baseline protocols have enough diversity of mechanism to forward data from the source to the destination, from different numbers of copies of data that they forward to varying sets of parameters that need to be considered for making the forwarding decision.

Indeed, the performance comparisons between those baseline routing protocols have been previously presented in various works. Yet, we aim to study their behavior in a unique use case where the current trend of the convergence between V2X communications and the Intelligent Transportation Systems (ITS) allows vehicles to be more involved in the smart city's data collection. Therefore, the main contributions of this paper are as follows.

\begin{enumerate}[nolistsep]
    \item Most of the previous works consider the forwarding of data between moving nodes, e.g., from vehicles to vehicles. In contrast, we compare them for a use case of the data collection from stationary connected-objects to a central server in smart cities. In the vision of convergence between V2X communications and the ITS infrastructures, different intermediary nodes with varying mobility profiles exist to assist in the forwarding process.  In this scenario, we discover several unique dynamics that show each protocol's strengths and weaknesses that need to be highlighted further.
    \item We investigate the performance of baseline routing protocols in unique data collection settings in smart cities, where multiple ITS infrastructure units such as RSUs can function as Internet Point-of-Presences. Public transport, such as public buses with their predetermined mobility, are also involved in the data collection. The results strengthen our belief that instead of designing complex routing protocols, we can develop a simple protocol that can be efficient and high-performing. We believe in assessing the strengths and weaknesses of those baseline routing protocols as a step towards the new routing protocol.
\end{enumerate}
\vspace{10pt}

Our study aims to show how such VDTN solutions are efficient for collecting sensors' data. To investigate each of the strategy's potential, we devise scenarios where a sufficient number of cars are equipped with communications and networking capabilities. We implement scenarios where 37 static wireless sensors are almost evenly spread out in the city. There are also 5 PoPs where vehicles can offload the collected data from sensors and forward them to the central server. We also explore the impact of different mobility patterns on the networking performance of the data collection system. Two types of vehicles: cars and buses are used in the simulation. Cars represent random vehicle movement, while buses moving along their predetermined route represent a predictable mobility pattern commonly found in smart cities. Ideally, the routing protocol used in this kind of setting needs to consider these different mobility types to exploit them and get better performances. Due to their simplicity, the four baseline routing protocols investigated in this study do not recognize these factors. Yet, it is crucial to understand how they perform under this unique scenario for future refinement. The Key Performance Indicators (KPIs) that we used for comparison are Delivery Probability, Average Latency, and Overhead Ratio, as commonly used in other previous works \cite{agussalim_comparison_2014, hadiwardoyo_deploying_2015, giannini_delay_2016, spaho_2016_evaluation, hernandez-jimenez_modeling_2019}.

We organized the rest of this paper as follows. Section II provides a brief history of vehicle-based data collection, an overview of the V2X Intelligent Transportation Systems (ITS) environment, and details on the baseline routing protocols for VDTN. Section III discusses some previous works already conducted within this subject and how our work differs. Section IV presents the performance evaluation, which includes the simulation setup and discussion of results. Lastly, Section V provides conclusions and possible future research.

\section{Background}

\subsection{Car-Based Data Collection: Early Days to Current Trends}

The idea of using vehicles for data collection has been around since the end of the last century. In sparse Wireless Sensor Networks (WSN), researchers have been proposing the use of mobile data collectors to reduce the power consumption of sensors, while at the same time achieving cost-effective connectivity. Data MULEs (Mobile Ubiquitous LAN Extensions) are mobile nodes that pick up data in one place and drop it elsewhere \cite{shah_data_2003}. With the three-tier architecture, such implementations aim to extend the network coverage and increase communications opportunities \cite{pereira2012delay}. A simulation study in \cite{anastasi_data_2008} shows an interesting direction to improve the energy efficiency of data collection with data mules, and another work in \cite{medjiah_sailing_2014} presents a Geographic Routing/Greedy Forwarding-based protocol for data mules in DTN.

As reported in the year 2000, a project called 'Roads towards the Future' was conducted by the Dutch ministry of transport and waterworks. The project introduced a field-trial of communication between vehicles and the infrastructure to exchange what they called the Floating Car Data (FCD) \cite{turksma_various_2000}. The data consist of the GPS-position of a sufficient number of vehicles frequently communicated to the central site via GSM links. The travel times of vehicles can then be measured accurately. The system's applications range from real-time data collection for traffic management to the compilation of very accurate Origin-Destination matrices complete with travel times. Currently, research towards various utilization of FCD are still continuing, as reported in \cite{astarita_floating_2020} and \cite{colombaroni_analysis_2020}, among others.      

Another generic in-vehicle data collection is the Probe Data. In \cite{uno_using_2009}, "Probe" is defined as a vehicle or person equipped with a GPS receiver. They propose an approach for evaluating the road network from travel time stability and reliability using probe data from a bus. More recent work in \cite{sekula_estimating_2018} combines vehicle probe data with machine learning techniques to estimate historical hourly traffic volumes. Meanwhile, another study in \cite{villiers_evaluation_2019} utilizes probe data for traffic management strategies for special events.

\subsection{The V2X Communications in The ITS Environment}

The term “smart” can be identified for future cities if they will be based on intelligent transportation solutions that embrace information and communication technologies \cite{pau_special_2019}. Subsequently, different stakeholders ranging from governmental agencies to automotive manufacturers have been significantly interested in vehicle-to-everything (V2X) communication as part of their development and deployment efforts for Intelligent Transportation Systems (ITS) \cite{arena_review_2020}. In the vision for smarter cities, vehicles can connect and exchange information with any devices. Ultimately, users, devices, and vehicles will form an ecosystem of Cooperative Intelligent Transportation Systems (C-ITS) \cite{silva_2017_opportunistic_chapter}.

Vehicle-to-Vehicle Communications (V2V) technology consists of wireless data communications between motor vehicles \cite{arena2019overview}, such as cars, taxis, buses, trucks, trams, and even trains. This communication's primary purpose is for accident mitigation, providing means for travelling vehicles to transfer data on their position and speed within an ad-hoc mesh network.  

Vehicle-to-Infrastructure Communications (V2I) communications allow interfacing between in transit vehicles with the road systems, such as traffic lights, street lights, lane markers, cameras, signage, and parking meters.  The standard V2I communications are wireless, bidirectional, and similar to V2V, utilizing Dedicated-Short-Range Communication (DSRC) frequencies to transfer data \cite{arena2019overview}.

Finally, Vehicle-to-Everything Communications (V2X) extends the V2V and V2I communications described above and represents generalization, where the data exchange takes place from vehicles to any entity they can interact with.  The vast technology covers other more specific types of communications, namely Vehicle-to-Pedestrian (V2P), Vehicle-to-Roadside (V2R), Vehicle-to-Device (V2D), and Vehicle-to-Grid (V2G) \cite{arena2019overview}. 

\subsection{The Baseline Routing Protocols for VDTN}
Here, we define four baseline VDTN routing protocols \cite{jain_routing_2004,vahdat_epidemic_2000,spyropoulos_spray_2005,pereira2012delay,benamar_routing_2014,abdelkader_performance_2016} that are studied in detail later.

\subsubsection{Direct Delivery Routing Protocol}
The Direct Delivery routing is a single copy forwarding approach at its simplest form, where a node that has the data forward it directly to its destination.  We can use this routing protocol performance as a benchmark that should be exceeded if more complexities are added to the designed routing strategy. Due to its simplicity, we can expect minimal network and buffer usage from this protocol.

\subsubsection{First Contact Routing Protocol}
The First Contact routing is a single copy forwarding approach where each node forwards messages randomly to the first node they encounter. This single copy of messages continues to hop randomly between in-range nodes until one reaches its destination. Nodes erase messages that they already relayed to another node, which means only a single copy of a message exists in the network. This strategy makes the routing protocol very efficient in occupying buffering spaces.   

\subsubsection{Epidemic Routing Protocol}
The Epidemic is a multiple copy forwarding approach where each node keeps copies of every message while also forwarding them to every other node they encounter until the messages reach the destination. Each node receives messages that they do not already have, with their buffering capacity as the only limitation. This approach ensures that at least one copy of each message will reach its destination in the earliest possible time, with the expense of flooding the networks with redundancy. As a benchmark, we can expect this routing as an upper limit in terms of network and buffer utilization. 

\subsubsection{Spray \& Wait Routing Protocol}
The Spray \& Wait routing is a more controlled multiple copy forwarding approach, where the number \textbf{\emph{L}} can be assigned to the protocol to specify the upper limit of copies that can be created per message by a node, as described in \cite{spyropoulos_spray_2005}. Furthermore, this routing protocol has two modes of operations: \emph{standard} and \emph{binary}.
In binary mode, if a node is the origin of the data, it will logically hold \textbf{\emph{L}} copies of the data specified initially (e.g., 6 is the default value). In standard mode, if a connection is established with another node that does not have a copy of the data, only a single copy is forwarded (the \emph{spray} phase), and \textbf{\emph{L-1}} copies continue to be held by the originating node. The originating node then can forward the remaining copies of the data to each node that it encounters next. The process continues until the originating node only hold the last copy of the data, when it stop forwarding copies to another node and only forward the data to its destination (the \emph{wait} phase). In binary mode, on the other hand, \textbf{\emph{L/2}} (rounded up) copies are forwarded each time the originating node meets another node without the data copy, while it will keep \textbf{\emph{L/2}} (rounded down) copies. As in the standard mode, the forwarding mechanism continues until the originating node has only a single copy of the data when it goes to the wait phase, which is basically similar to the direct delivery routing.

\subsection{The Key Performance Indicators (KPIs)}

The performances of DTN, and consecutively VDTN, can be measured by several Key Performance Indicators (KPIs). In this study, we utilize three commonly used KPIs in previous studies on this topic \cite{agussalim_comparison_2014, hadiwardoyo_deploying_2015, giannini_delay_2016, spaho_2016_evaluation, hernandez-jimenez_modeling_2019}. We discuss on each of them in the following.

\subsubsection{Delivery Probability}
The Delivery Probability (DP), or also referred to as the Delivery Ratio, is the total number of messages successfully delivered to their destination divided by the total number of generated messages at the originating nodes as defined in Equation \ref{eq1}.

\vspace{5pt}
\begin{equation}
    \label{eq1}
    \textrm{DP} = \frac{\textrm{Number of delivered messages}}{\textrm{Number of generated messages}}
\end{equation}
\vspace{5pt}

The ideal condition is for the maximum delivery probability value of 1, where all the generated messages are also successfully delivered to the destinations. 

\subsubsection{Average Latency}
Also referred to as the Average Delivery Delay, the Average Latency (AL) is the average time it takes from the messages' creation at the sources to the time they are successfully delivered at the destination, as defined in Equation \ref{eq2}.

\vspace{5pt}
\begin{equation}
    \label{eq2}
    \begin{split}
    \textrm{AL = }
    \frac{\sum_{n=1}^{N} \textrm{Message arrival time} -\textrm{Message creation time}}{\textrm{\emph{N} delivered messages}}, \text{if } N\geq 1
    \\ \emph{cannot be defined}, \text{if } N = 0
    \end{split}
\end{equation}
\vspace{5pt}

The desired condition is for messages to instantaneously reach their destination, i.e., the average latency of close to zero, particularly for critical applications. Yet, in the intermittent connectivity of vehicular networks, the ideal condition of very low latency is almost impossible to achieve. Therefore, they are more suitable for delay-tolerant applications, whereas data with an average latency in the order of seconds, minutes, or even hours, can still be useful. Nevertheless, the goal is to keep the average latency as low as possible.     

\subsubsection{Overhead Ratio}
The Overhead Ratio (OR) shows the ratio of the total number of transmitted messages in the entire network compared with the total number of delivered messages, as shown in Equation \ref{eq3}.

\vspace{5pt}
\begin{equation}
    \label{eq3}
    \begin{split}
    \textrm{OR = } 
    \frac{\textrm{\emph{T} transmitted messages}-\textrm{\emph{N} delivered messages}}{\textrm{\emph{N} delivered messages}}, \text{if } N\geq 1 
    \\ \emph{cannot be defined}, \text{if } N = 0
    \end{split}
\end{equation}
\vspace{5pt}

The ideal value for the Overhead Ratio is 0, which happens only when the source of a message or messages directly deliver them to their receiver. If intermediary nodes are involved in the delivery, relaying messages through those intermediaries is considered as excesses or overheads. Hence, relaying copies of messages that do not eventually reach the destination will also be counted as overheads, including copies being rejected by the receiver. In most cases, the receiver only accepts a unique message once, i.e., the first time it arrives. The parameter was also referred to as the Network Overhead Ratio in some references because it directly affects the network's resource usage, such as energy consumption for processing and communications, as well as bandwidth allocation. A low Network Overhead Ratio is a vital characteristic of an efficient and scalable data collection system.

\begin{figure}[h!]
\centering
\includegraphics[width=0.87\linewidth]{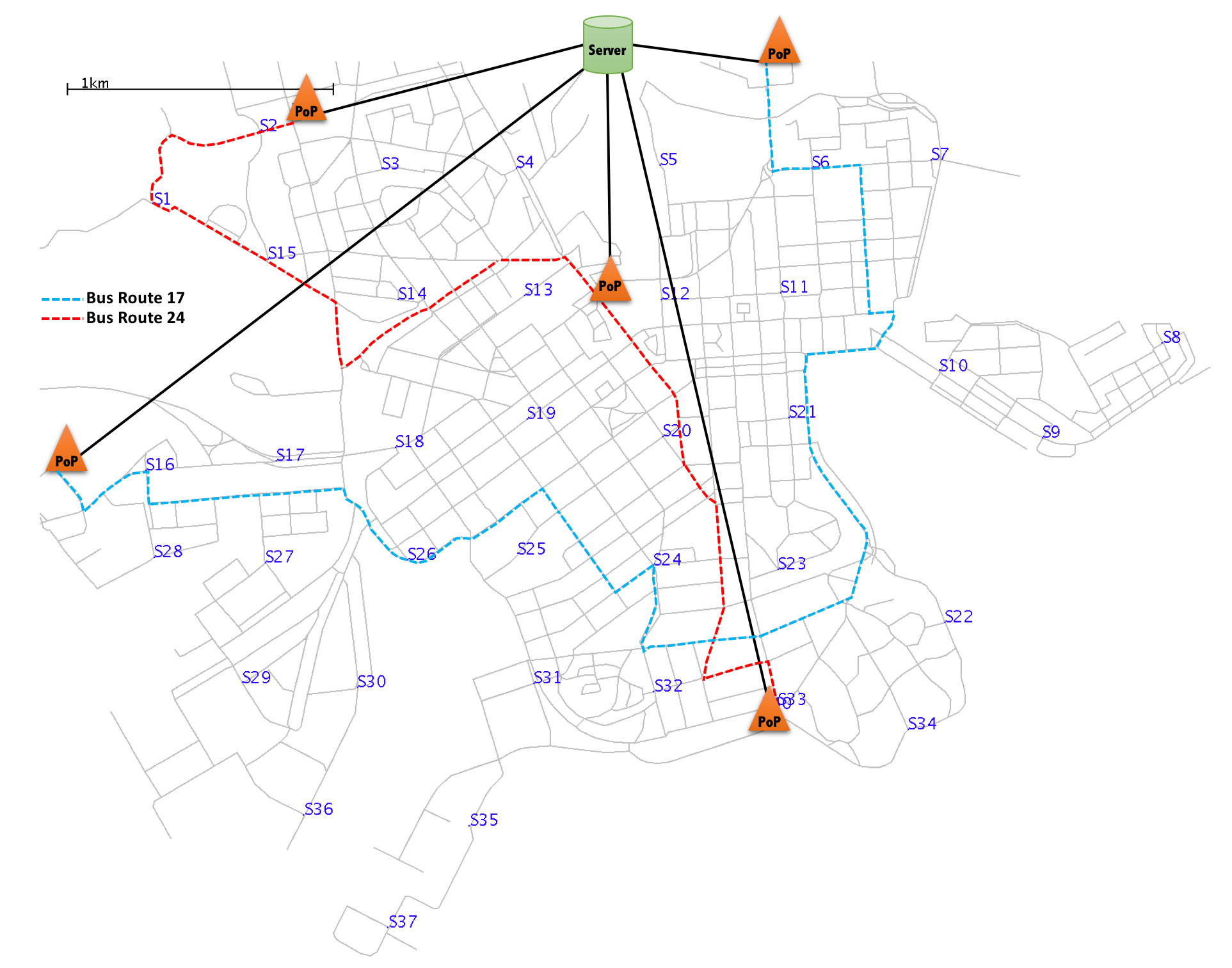}
\vspace{.7em}
\caption{The Simulation Map and Configuration}
\label{fig3}
\end{figure}

\section{Performance Evaluation}

The Opportunistic Networking Environment (ONE) simulator \cite{keranen_one_2009} was used to evaluate the VDTN-based data collections' networking performances. As we mentioned previously, we apply three KPIs for the performance evaluation, namely Delivery Probability, Average Latency, and Overhead Ratio \cite{agussalim_comparison_2014, hadiwardoyo_deploying_2015, giannini_delay_2016, spaho_2016_evaluation, hernandez-jimenez_modeling_2019}.

We utilize a route-based movement for buses and random-waypoints movement for cars as mobility models in a real-world city map in the simulation. In this paper, our main goals are to have the performance comparisons between the baseline routing protocols and understand the trends as the density of the vehicular networks increases, which can be achieved even with these simplest mobility models. As discussed in \cite{dede_simulating_2018}, even though trace-based mobility models give realistic movement patterns, a complete set of traces for the newly investigated scenario, such as in our case, may not be available. Moreover, they produce a static scenario where the number of nodes stays the same, limiting our possibility of observing the effect of a growing vehicular network. However, we are considering to conduct simulations using suitable location-based-traces for evaluations in our future work. 

\subsection{Simulation Setup}

The city of Helsinki’s map, a standard map in the simulator, with a land area size of approximately nine square kilometers, was used to set up the city data collection scenario. We proposed a delay-tolerant environmental monitoring system that acquires data from 37 wireless sensors which spaced almost evenly every 500 meters to each other in the city. Data from those stationary wireless sensors in the environment need to reach one of 5 available Internet Point-of-Presences (PoPs) strategically placed along the bus routes. The mapping of the simulation is shown in Figure ~\ref{fig3}. We assume the wireless sensor to be limited in its power capability, i.e., battery-powered or energy harvesting. Therefore, it can only perform a basic data forwarding mechanism, such as the First-Contact routing, to transmit data only to the first in-range vehicle. 

With several sets of simulation scenarios, we increase the number of cars involved in the data collection, while the number of buses remain the same. The cars involved are from 3 to 90, and 2 buses for each route make up the total of 4 buses assisting in the data collection. One of the PoP is positioned at the city center where most of the traffic converged, while four PoPs are positioned at each end of the two bus routes.
In this configuration, a bus will pass at least two PoPs as they move along their route, while there are no guarantee that cars will encounter  one. On the other hand, the benefit of utilizing cars is that they can go to various places in the city and opportunistically pick up data from in-range sensors, which is not the case for buses.
With the understanding of the trade-off, we aim to observe some baseline routing protocols' networking performance in this setting and find the gap for future refinement. 

Each sensor has a ZigBee wireless link profile with a 10 m communications range and data rate of 250 kbps. We choose 10 m as a conservative value for ZigBee's communication range to include the possibility of Non-Line-of-Sight (NLOS) signal propagation condition between sensors and vehicles and even mimic indoor sensor placements. The sensors generate 10 Bytes of data every 5 minutes over the first 7 hours of the 12 hours total simulation duration. The data have a Time to Live (TTL) of 5 hours. The 10 Bytes of data is for a single environmental data measurement described in \cite{kang_public_2016}. We choose the ZigBee communication technology as an archetype of a short-range, low-power sensor node operating on battery for months, even years. Each car, which can be connected wirelessly to the sensor via their identical ZigBee link profile, has a pseudo-random initial placement and mobility. For its V2X communications, the car also has an ITS-G5 wireless link profile, with a 300 m communications range and data rate of 6 Mbps. The PoPs have an identical ITS-G5 connection with the cars for data retrieval, and each also has a direct link to the destination server. We implement 5 MB of buffer size for cars and buses, which enable unconstrained evaluation of the networking performance. The number of cars in the scenario varied from 3 to 90, representing an average cars density of 0.33 to 10/km\textsuperscript{2}. All parameters and values used in the simulation are provided in Appendix A of this paper.

\begin{figure}[!hbt]
\centering
\includegraphics[width=0.93\linewidth]{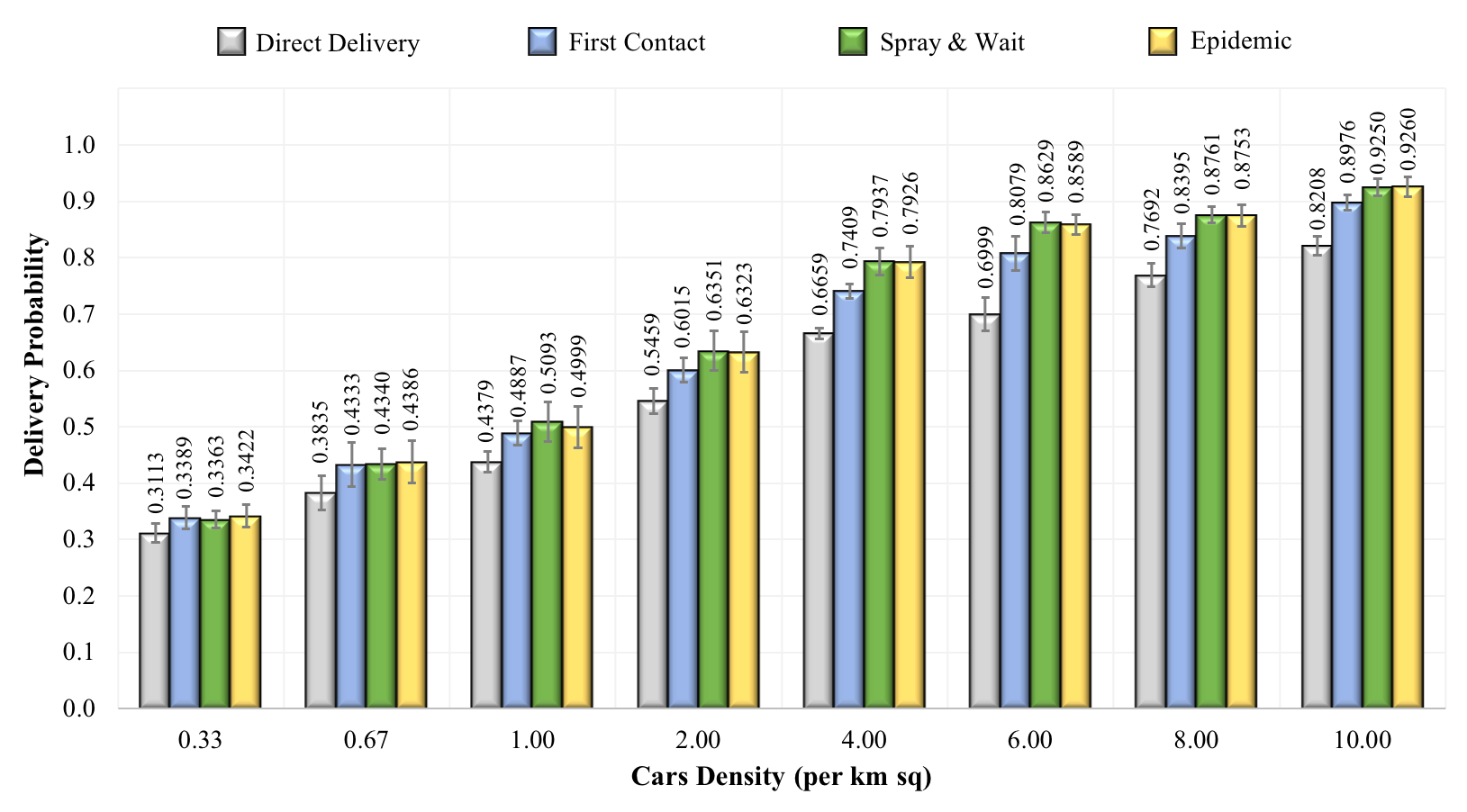}
\vspace{.7em}
\caption{The Probability of Delivery Comparison}
\label{fig4}
\end{figure}

\subsection{Results and Discussion}

In all the scenarios, the first in-range vehicle (car or bus) to the sensor will pick-up the available data. The vehicle then carries and forwards the data based on the routing protocol that they implement. We need to remember that the cars' possibility to reach the PoP is uncertain due to its random mobility setting, while the buses will always encounter at least 2 PoPs along their route. In terms of delivery probability, data carried by buses are desirable, but it may experience a longer waiting time at the sensor, which increases its latency. On the other hand, cars may pick up data from the sensor more often, as statistically more cars will be available in the city. But if the data is only carried by cars, then the delivery probability will suffer, as there is the uncertainty whether it will reach one of the PoP.

Furthermore, we need to emphasize on the dynamic proportion of vehicles involved in the data collection, as it crucially relates to the KPI results' interpretation. In our scenario, buses are considered the mobile network's backbone, as PoPs are placed along their routes. A constant number of buses, 4 in total, are deployed in all scenarios, while the number of cars increases from 3 to 90. Hence, more buses are involved in the network in the scenario with the least number of cars. The proportion of cars in the network then gradually increases while the number of buses remains the same, changing the ratio between vehicles with random and predetermined mobility. The effects of this dynamic, combined with the different mechanisms that each of the baseline routing protocol deploys, are explained in the following discussion.

We first evaluate each routing protocol's delivery probability and show their comparison in Figure ~\ref{fig4}. It shows that when the cars density was low, i.e., sparse network, the two single-copy approaches, the First Contact and the Direct Delivery, achieved comparable delivery probability to the two multi-copy routing protocols, the Spray \& Wait and the Epidemic. But as the cars density increases, i.e., the network becomes denser, their delivery probability begins to lag. Initially, we expected that the Epidemic routing protocol would always outperform the other three routings with its maximum copies approach. Yet, surprisingly, the Spray \& Wait's delivery probabilities are comparable to the Epidemic's in all network densities. In the larger part, it was because most of the undelivered messages were never transferred from sensors to in-range vehicles until they ran out of TTL and dropped from the sensor's buffer. In a smaller portion, the cause was the unsuccessful messages transfer from vehicles. It turned out that in the intermittent connection with moving vehicles (vehicles with vehicles or vehicles with stationary nodes), the limited contact duration and the number of messages queuing for transmission in the buffer diminished the advantage of having maximum copies of messages in the network. It is possible that when vehicles' buffer becomes flooded by Epidemic messages: there is a lower probability that a particular message can be transferred in a brief window of contact between nodes (sensors, cars, buses, and PoPs). Moreover, the faster the vehicle move, the narrower the transferring window becomes, as previously studied in \cite{er_contact_strategy_2018}. The dynamic highlighted that flooding the network with unlimited data copies seems to give minor gain, if not none, in terms of delivery probability in our scenario. 

In the Direct Delivery routing, vehicles will keep all data they received from sensors and only do the forwarding if one of the PoP is in the communication range. Therefore, in our scenario, only data picked up by buses will have guaranteed delivery due to the PoPs placement, while data collected by cars will have none of the certainty. It explains why this routing protocol's delivery probabilities are the lowest, and even more so as the cars density increases.

On the other hand, in the First Contact routing, the vehicle with the data will always forward them to the first node (car, bus, or PoP) that it encounters. In this approach, two mechanisms with opposite effects can take place in the data collection. The first one is when the car forwards the data to an in-range bus, i.e., a condition of guaranteed delivery, which is desirable. The second one is when the bus needs to 'give up' its data to a nearby car, i.e., cancellation of guaranteed delivery, which is undesirable. Yet, despite this challenging dynamic, the results show that the First Contact performs better than Direct Delivery in its probability of delivery. It turned out that data picked up by cars from sensors will have a better chance of reaching one of the PoP if it is opportunistically forwarded to the next in-range vehicle rather than to being kept for direct delivery. The advantage becomes more significant if the car is stationary for a long time; offloading the data would increase its delivery chance.  

In the Spray \& Wait and Epidemic routing, with their multi-copies approach, the vehicle with the data will forward them to all in-range nodes while also keeping the original for direct delivery to the destination. It resulted in a higher delivery probability as more paths are possible for the data to reach the PoPs. The results also give one crucial point: the Spray \& Wait's Delivery Probabilities are comparable to those of the Epidemic's, hinting to the advantage of forwarding limited copies and the inefficiency of deploying maximum copies in the scenario.    

\begin{figure}[!h]
\centering
\includegraphics[width=0.93\linewidth]{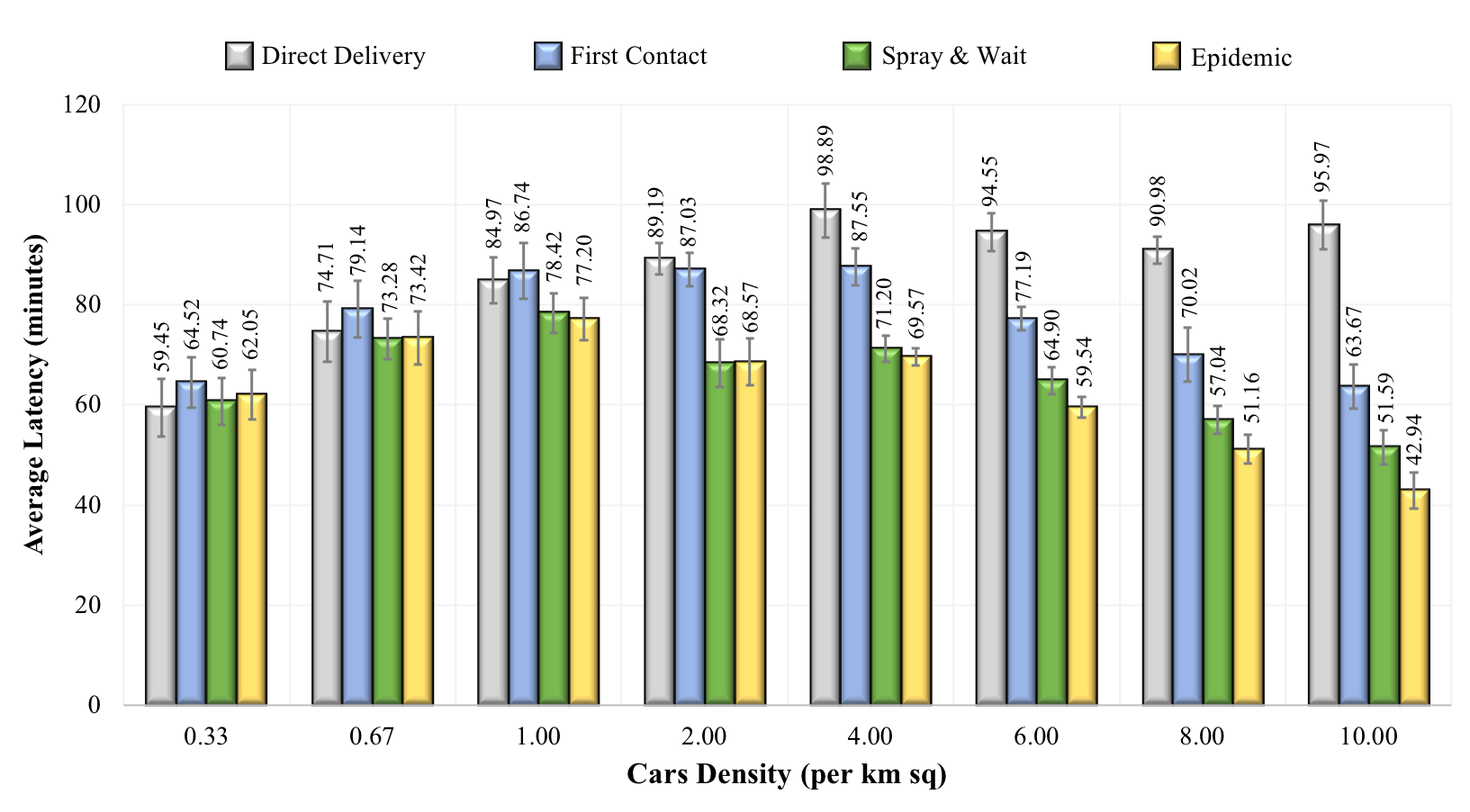}
\vspace{.7em}
\caption{The Average Latency Comparison}
\label{fig5}
\end{figure}

Figure ~\ref{fig5} compares the routing protocols in terms of the average latency. It shows two opposite trends for majority of the routing protocols in relation to the available number of cars in the network. The Direct Delivery routing protocol is the only exception to the trend. 

For the Direct Delivery routing, increases in the cars density are also accompanied by increases in the average latency until it reaches around 99 minutes at a cars density of 4 per km\textsuperscript{2}. At that point, the average latency seems to come to its maximum value, where it remains below 99 minutes, even with the increase in the cars density. In the minimum car density of 0.33 per km\textsuperscript{2}, Direct Delivery's average latency is also the lowest at 59.45 minutes. In interpreting this result, we need to relate to their low delivery probability shown in Figure ~\ref{fig4}, of around 0.31 only, where the average latency is calculated only from the successfully delivered data. There are only 3 cars in the entire network in this cars density, less than the number of buses, which is 4. Therefore, there is a high probability that the delivered data were previously picked up by buses, which then carry them to one of the PoPs along their route, causing the lower average latency. As the cars density and the delivery probability increase, more cars pick the data instead of buses. Consequently, the average latency is also going up due to data being kept longer in cars with the Direct Delivery routing protocol. The results also suggest a particular maximum value of average latency exists for the data collection with this routing protocol, although it will undoubtedly depend on the car's mobility. Yet, with about 0.82 delivery probability at the highest cars density in our simulation, there is merit for its implementation in denser networks for applications that can tolerate a longer latency and an incomplete set of time-series data.  

For the First Contact routing, although it has a different forwarding mechanism than the Direct Delivery, the average latency trend is quite similar at the lower cars density, where increases in the density also lead to increases in the average latency due to the same mechanism. The difference is that the trend reverses at a cars density of 6 per km\textsuperscript{2} when the average latency starts to decrease with an increase in the cars density. One important note in the minimum cars density of 0.33 per km\textsuperscript{2}; the First Contact average latency of 64.52 minutes is higher than the Direct Delivery. The higher average latency might be due to the undesirable effect of buses 'giving up' its data to in-range cars, which we explained earlier. A more advantageous mechanism would be for buses to keep their data until they reach one of the PoPs, a gap that can be refined for a better routing protocol.

The two multi-copies routing protocols, the Spray
\& Wait and Epidemic, also possess a similar trend to the First Contact routing. They differ in that they have lower average latency and the reverse in the trend for them starts earlier at a cars density of 2 per km\textsuperscript{2}. Figure ~\ref{fig5} also shows that the Spray \& Wait only lagging the Epidemic slightly in terms of its average latency, particularly at the lower cars density, even though it only deploys up to 6 copies of each data in the forwarding process. It is a crucial advantage that will be highlighted further when discussing its Overhead Ratio comparison with the Epidemic later on.   

If we observe Figure ~\ref{fig5} further and compare the Average Latency of the First Contact and Spray \& Wait, we can find that the First Contact's Average Latency is always higher than Spray \& Wait's in all of the cars densities. It is primarily due to the higher number of data copies that Spray \& Wait allowed to forward, which increases the possibility that one of the copy will reach a PoP sooner.

Interestingly, the dynamic is also influenced by the inclusion of buses in the data collection network and how the Spray \& Wait routing 'accidentally' harnesses their potential. At a specific condition where a bus received a copy or copies of the data, the two routing protocols behave differently. When a bus with First Contact routing received a single copy of data, there is no guarantee that it will keep the data until it reaches one of the PoPs. As we pointed out earlier, there is the possibility that the bus will have to forward the data to an in-range car before it encounters any PoPs. Contrary, when a bus with Spray \& Wait routing received a copy or copies of data, it will at least keep a copy for direct delivery, consequently elevating the delivery probability and potentially minimizing the average latency. The 'accidental' behavior is a strength that needs to be embedded into a better routing strategy for the specific data collection purpose.

\begin{figure}[!h]
\centering
\includegraphics[width=0.93\linewidth]{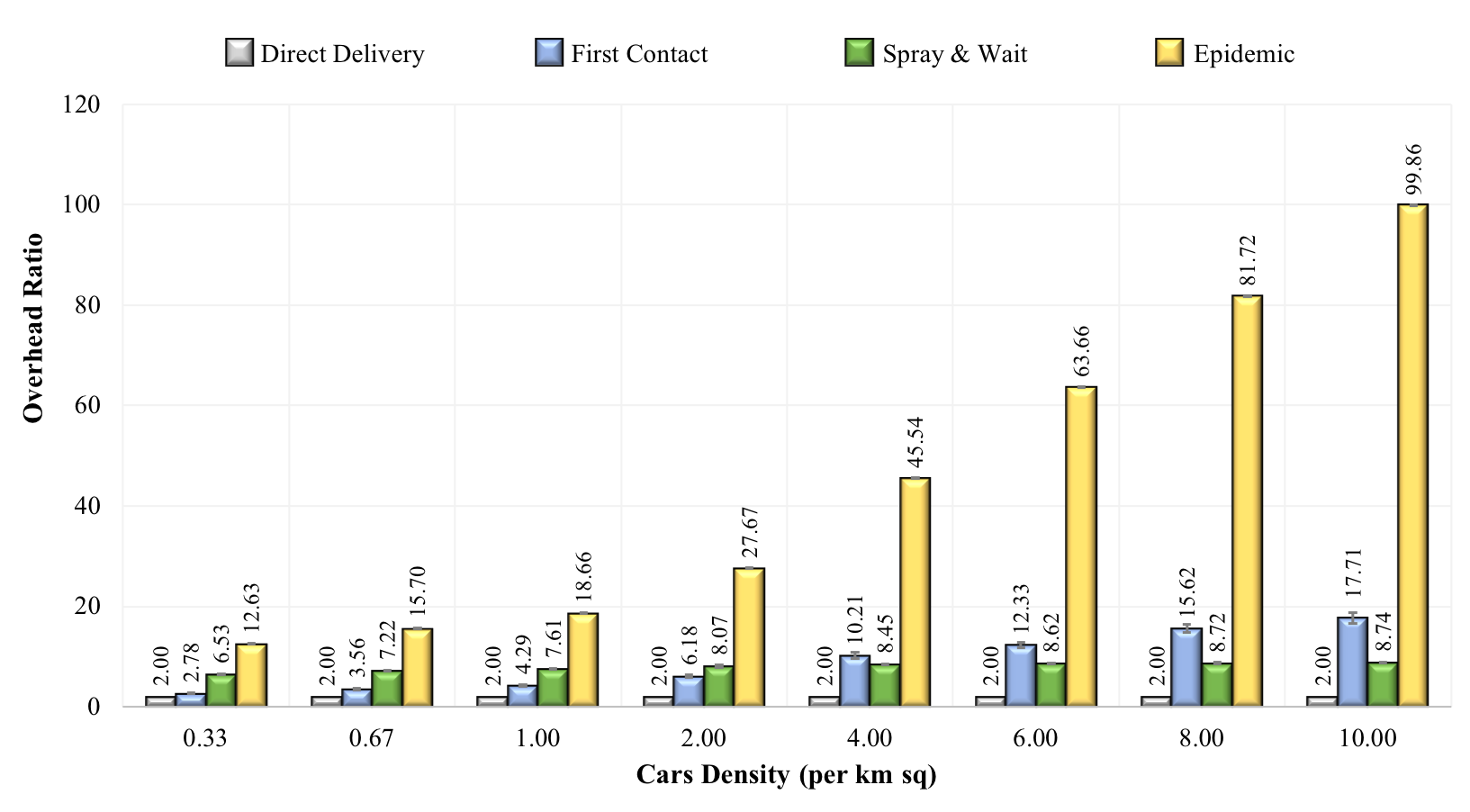}
\vspace{.7em}
\caption{The Overhead Ratio Comparison}
\label{fig6}
\end{figure}

Figure ~\ref{fig6}, on the other hand, captures each routing protocol's networking cost in terms of their overhead ratio. It shows contrasts between each routing protocol in the single-copy and the multi-copies groups. In the single-copy group, the Direct Delivery routing produces the lowest overhead ratio in all cars densities, a constant value of 2, as the single-copy data forwarding only happens from sensors to cars or buses, and then from cars or buses to one of the PoPs. On the contrary, the First Contact routing showed an increase of overhead ratio as the network becomes denser. The strategy of always forwarding the data to the next vehicle in-range makes the overhead ratio goes up as more vehicles become available to receive the data. There is no restriction on how many data forwarding processes can take place, until eventually the data reach one of the PoPs. There is also no distinction on which vehicles to forward the data to, which make the data transfer from buses to cars also possible. As described earlier, it would be more efficient if there is a distinction on which kind of vehicles to forward to, e.g., buses which have higher delivery probability should not forward data to cars. The added logic will lead to lower average latency and overhead ratio, as well as higher delivery probability.

We can also observe both similarity and disparity of the Overhead Ratio trend in the multi-copies routing pair. Spray \& Wait and Epidemic both experience an increase in the Overhead Ratio as the number of cars grows in the network. The difference is that the Spray \& Wait's Overhead Ratio seems to reach its saturation value of around 8 at a cars density of 2 per km\textsuperscript{2} and only has minor increases as the cars density goes up. The dynamic occurs because the Spray \& Wait routing implements a limited number of copies and, consequently, a limited number of forwarding strategy. Vehicles only forward a limited number of data copies to other vehicles. Each vehicle then does a direct delivery to one of the PoPs when only a single copy of the data remains in their possession. Therefore, the so-called 'wait' mode limits the communication hops and consequently reduces the Overhead Ratio. The opposite happens to the Epidemic, where its Overhead Ratio goes up sharply as the cars density increases. The Epidemic routing deploys unlimited data copies and communication hops strategy at the expense of inundating the network with more overheads, even though it manages to achieve the lowest Average Latency in a denser network. Indeed, the trade-off will be a concern to its scalability in an ever-growing network and further emphasizes the advantage of Spray \& Wait's limited-copies strategy.

Finally, from Figure ~\ref{fig6}, another revealing dynamic can be observed when we compare the Overhead Ratio of First Contact and Spray \& Wait routing. In a sparse vehicular network, i.e., cars density of 0.33 to 2 per km\textsuperscript{2} in our scenario, First Contact's Overhead Ratios are lower than the Spray \& Wait's. While in a denser network, First Contact's Overhead Ratios surpass Spray \& Wait's, which only increases slightly. The First Contact does not limit the number of forwarding (hops) for each data until one of the PoPs is reachable. There are limited vehicles at a lower cars density to forward the data. Only a single-copy is forwarded when one becomes in range, leading to a lower Overhead Ratio. Contrary, more vehicles are available in a denser vehicular network to forward the single-copy of data, for which the increasing number of hops give rise to the Overhead Ratio. 

The Spray \& Wait, on the other hand, starts with 6 copies of each data available to immediately forwarded to in-range vehicles. Thus, even when the number of cars is low, each vehicle can forward specific data to more than one vehicle in their `spraying' phase, causing a higher Overhead Ratio compared to the First Contact. A reversal happens as the cars density increases: vehicles with Spray \& Wait strategy forward a limited number of data copies which constrains the number of hops, even though plenty of vehicles are available to receive data from them. As we might recall, vehicles with this routing strategy can only forward copies until they are left with the last copy of each data, when they stop `spraying' and start their `wait' phase to deliver data to one of the PoPs directly. The limited hops dynamic resulted in Spray \& Wait's lower Overhead Ratio compared to First Contact's in denser networks.

\vspace{5pt}
\begin{table*}[h!]
\setlength{\arrayrulewidth}{0.3mm}
\centering
\renewcommand{\arraystretch}{1.0}
\caption{Summary of The Baseline Routing Protocols' Strengths and Weaknesses}
\vspace{.7em}
\footnotesize
\begin{tabular}{|p{2.5cm} |p{5.6cm} |p{5.6cm}|}
\hline
\textbf{Routing Protocol}   
        & \textbf{Strengths}    
        & \textbf{Weaknesses}    \\
\hline

\textbf{Direct Delivery}     
& * Fairly high Delivery Probability and showing an upper-limit of Average Latency in a denser network; might be applicable for certain applications. \newline
* Constant Overhead Ratio and the lowest among the baseline routings, offering the most efficient network usage.
& * Lowest Delivery Probability and highest Average Latency among the baseline routings. \\
\hline

\textbf{First Contact}       
& * Fairly high Delivery Probability and shows a decreasing Average Latency as the network become denser. 
& * Continuous increases in Overhead Ratio as the network become denser. \newline
* It does not have a mechanism to recognize vehicles with predetermined mobility; therefore, the single-copy of data might be offloaded to cars that have a lower probability of delivering them.\\
\hline

\textbf{Spray \& Wait}       
& * Tied-highest Delivery Probability (with Epidemic) among the baseline routings and shows a decreasing Average Latency as the network become denser. \newline
* Moderate Overhead Ratio with minor increases as the network grows. \newline
* The \emph{wait} phase, when a forwarding vehicle keeps at least a copy of the data, 'accidentally' ensures the direct delivery of the data by vehicles with predetermined movement.    
& * It does not have a mechanism to recognize vehicles with predetermined mobility; therefore, data that is already relayed to vehicles with predetermined mobility continue to be forwarded to cars with a lower probability of delivering them, consequently increasing the Overhead Ratio.\\
\hline

\textbf{Epidemic}            
& * Tied-highest Delivery Probability (with Spray \& Wait) and the lowest Average Latency among the baseline routings.
& * Extremely high Overhead Ratio, which will continue to increase in a growing network.\\
\hline

\end{tabular}
\label{tabA2}
\end{table*}

To summarize, Table ~\ref{tabA2} shows the strengths and weaknesses of the four baseline routing protocols when applied to the proposed data collection scheme. All the results above point out that the baseline routing protocols evaluated in this study do not have a mechanism to exploit the full potential of having vehicles with predetermined mobility, such as buses, in the data collection process. Therefore, the advantage of having ITS infrastructures to assist in the system is also overlooked. Our performance evaluation indicates the need to develop a routing protocol that recognizes and utilizes the advantage of different kinds of mobility in the field for data collection, such as the work on a hierarchical VDTN routing protocol named DC4LED presented in \cite{er_dc4led_2019} and \cite{er_CityView_2019}. We believe that for the specific purpose of data collection that we already discussed, a simple and efficient routing protocol can be deployed for the delay-tolerant data delivery, contrary to the implementation of optimized solutions that might be resource-demanding and difficult to standardize.

\section{Conclusion and Future Works}

This paper presented the performance comparison of four baseline VDTN routing protocols in a unique ITS-assisted data collection setting in smart cities. In the VDTN-based data collection scenario, several Internet Point-of-Presences are available and strategically located in the city to assist in the data delivery from connected-objects to the application server. We assessed the four routing protocols in this specific scenario to see gaps that can be refined. These protocols have sufficient diversity of mechanisms to forward data, giving plenty of insight into their impact on the Key Performance Indicators. Furthermore, the inclusion of two types of vehicles, each with their distinct mobility pattern, into the simulation also reveals some crucial forwarding dynamics.

The comparison between the baseline routing protocols showed that generally multi-copy strategies perform better by offering higher delivery probability and lower average latency. Their drawback was on the high overhead ratio that will burden the network in a dense environment such as smart cities. Moreover, we also observed different trends in varying vehicles density, wherein a sparse network the single-copy approaches actually have a comparable performance with the multi-copy strategy. It was only when the network became denser that the advantage of forwarding limited multi-copies data became apparent, suggesting implementing a different routing strategy based on the vehicular network density. Our work also emphasizes some gaps in the routing strategies that can be refined for better performances. 

Our future works will include refining the DC4LED routing protocol to incorporate the advantage of the limited-copies approach highlighted by the Spray \& Wait routing performances in this paper. The DC4LED routing protocol already consists of a mechanism that recognizes and utilizes specific mobility patterns from various vehicles available in the city. It forwards data hierarchically to maintain efficiency and scalability. The protocol will also need to be tested in a more realistic vehicle's mobility pattern provided by real-world GPS traces to increase the KPI values' accuracy. 

\section*{Acknowledgements}
This research was supported by a Doctoral Scholarship from the Indonesian Endowment Fund for Education (LPDP) of the Indonesia's Ministry of Finance, under Contract Number PRJ-43/LPDP.3/2016.

\clearpage



\appendix
\section*{APPENDIX A. Parameters and Values in The Simulation}

\begin{table}[h!]
\setlength{\arrayrulewidth}{0.4mm}
\centering
\renewcommand{\arraystretch}{1.0}
\footnotesize
\begin{tabular}{|p{6.0cm} p{5.0cm}|}
\hline
\textbf{Parameters}         & \textbf{Values}    \\
\hline\hline
Map size                    & 4.5 km x 3.4 km       \\
\hspace{3mm}Land area       & approximately 9 km$^2$   \\
Simulation time             & 12 hours              \\
Simulation warm up time     & 200 s                 \\
Message generation window   & 7 hours               \\
Message Time-to-Live (TTL)  & 5 hours               \\
Messages created by sensors & 3071                  \\
\hline\hline

\textbf{Sensors}                 &              \\
\hline
\hspace{3mm}Number of sensors    & 37                \\
\hspace{3mm}Movement model      & Stationary        \\
\hspace{3mm}Message size        & 10 B              \\
\hspace{3mm}Message generation interval & 5 minutes \\
\hspace{3mm}Buffer size         & 64 KB             \\
\hline
\hspace{3mm}Interface type      & ZigBee link profile            \\
\hspace{3mm}Transmission range      & 10 m          \\
\hspace{3mm}Transmission rate       & 250 Kbps      \\
\hline\hline

\textbf{Cars}                   &                       \\
\hline
\hspace{3mm}Number of cars      & 3; 6; 9; 18; .. 72; 90 (corresponds to cars density of 0.33 to 10 per km$^2$) \\   
\hspace{3mm}Movement model      & Random Waypoints \& Shortest-path Map-based   \\
\hspace{3mm}Movement speed      & 10 - 50 km/h          \\
\hspace{3mm}Stationary time at  &                       \\
\hspace{3mm}waypoints           & between 1 - 120 min. \\
\hspace{3mm}Buffer size         & 5 MB                      \\
\hline
\hspace{3mm}Interface\#1 type   & ZigBee link profile  \\
\hspace{3mm}Transmission range  & 10 m              \\
\hspace{3mm}Transmission rate   & 250 kbps          \\
\hline
\hspace{3mm}Interface\#2 type   & ITS-G5 V2V link profile \\
\hspace{3mm}Transmission range  & 300 m              \\
\hspace{3mm}Transmission rate   & 6 Mbps         \\
\hline\hline

\textbf{Buses}                      &                       \\
\hline
\hspace{3mm}Number of buses         &  4 (2 for each route)   \\
\hspace{3mm}Movement model          & Fixed Waypoints \& Shortest-path Map-based   \\
\hspace{3mm}Movement speed          & 10 - 30 km/h          \\
\hspace{3mm}Stationary time at      &               \\
\hspace{3mm}waypoints               & between 10 - 20 sec. \\
\hspace{3mm}Buffer size             & 25 MB                    \\
\hline
\hspace{3mm}Interface\#1 type   & ZigBee link profile  \\
\hspace{3mm}Transmission range  & 10 m              \\
\hspace{3mm}Transmission rate   & 250 kbps (Zigbee)           \\
\hline
\hspace{3mm}Interface\#2 type   & ITS-G5 V2V link profile \\
\hspace{3mm}Transmission range  & 300 m              \\
\hspace{3mm}Transmission rate   & 6 Mbps         \\
\hline\hline

\textbf{Internet Point-of- Presence (PoP)}    & \\
\hline
\hspace{3mm}Movement model      & Stationary   \\
\hspace{3mm}Buffer size         & 100 MB             \\
\hline
\hspace{3mm}Interface\#1 type   & ITS-G5 V2I link profile \\
\hspace{3mm}Transmission range  & 300 m              \\
\hspace{3mm}Transmission rate   & 6 Mbps         \\
\hline

\end{tabular}
\label{tab4.2}
\end{table}

\clearpage

\renewcommand*{\bibfont}{\normalfont}
\printbibliography[title=REFERENCES]


\clearpage

\section*{BIOGRAPHIES OF AUTHORS} 
\vspace{-.7em} 

\begin{biography}[{\includegraphics[width=2.5cm,height=4cm,clip,keepaspectratio]{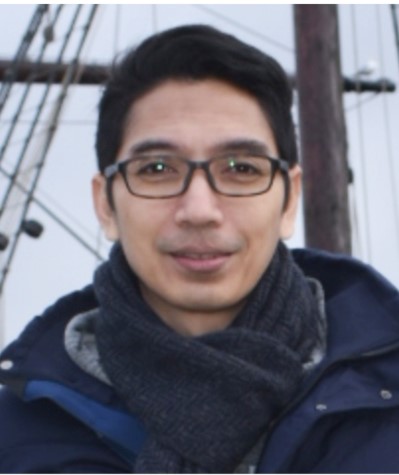}}]
\textbf{Ngurah Indra Er} 
is a lecturer in the Department of Electrical and Computer Engineering, Udayana University (UNUD), Bali, Indonesia, since 2002. He completed his PhD from IMT Atlantique, Rennes, France in 2021. He obtained his M.Sc. degrees in Communications Engineering from University of Birmingham, U.K., in 1999 and his B.Eng. in Electrical Engineering from Sepuluh Nopember Institute of Technology (ITS) Surabaya, Indonesia, a year earlier. His current research interests include Internet of Things, Opportunistic Networks, Vehicular Networks, and Smart City Data Collection.
\end{biography}

\begin{biography}[{\includegraphics[width=2.5cm,height=4cm,clip,keepaspectratio]{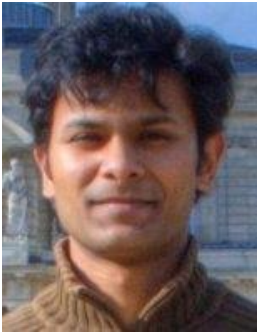}}]
\textbf{Kamal Deep Singh} is currently an Associate Professor at
Telecom Saint Etienne / University Jean Monnet, France. He received the B.Tech. degree in Electrical Engineering from Indian Institute of Technology (IITD), Delhi, India in 2002 and obtained his Ph.D. degree in computer science from University Rennes 1, France in 2007. He then worked as a Postdoc researcher in the Dionysos group at INRIA and at Telecom Bretagne, Rennes, France, where he developed many components of QoE estimation tools and worked on the analysis of video-based applications. His current research interests include Quality of Experience (QoE), Internet of Things, Complex Event Processing and Big Data.
\end{biography}

\begin{biography}[{\includegraphics[width=2.5cm,height=4cm,clip,keepaspectratio]{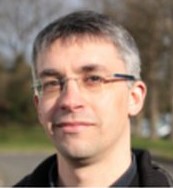}}]
\textbf{Christophe Couturier} 
joined IMT Atlantique, France in 2014 where he holds a position of Assistant Professor. After an engineering degree in electronics at INSA of Rennes in 2001, he worked for 13 years as a technical manager for innovative telecommunication projects in the field of transportation (rail and road). Currently, his main areas of interest are Intelligent Transportation Systems (ITS) and Mobile Networks. \end{biography}

\begin{biography}[{\includegraphics[width=2.5cm,height=4cm,clip,keepaspectratio]{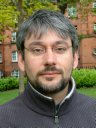}}]
\textbf{Jean-Marie Bonnin} 
is currently a Professor (HDR) in Networks and Computer Science at IMT Atlantique, France and the leader of the Inria/EASE research team. After a PhD degree in computer science at the university of Strasbourg in 1998, he came to ENST Bretagne for a research position. Since 2001, He has been mainly interested in the convergence between IP networks and mobile telephony networks, and therefore in heterogeneous handover management. More recently, he has been involved in projects dealing with the mobility of the networks and its application to Intelligent Transportation System (ITS). His current research interest is on how to provide pervasive applications with a good perception of their environment through localized interactions, specifically in cooperative autonomous vehicles. He is also a co-founder and scientific advisor of the YoGoKo startup which provides communication solutions for cooperative ITS including autonomous vehicles. 
\end{biography}

\end{document}